\documentclass[%
 reprint,
showpacs,
preprintnumbers,
 amsmath,amssymb,
 aps,
]{revtex4-1}

\usepackage{graphicx}
\usepackage{dcolumn}
\usepackage{bm}
\usepackage[T1]{fontenc}
\usepackage{textcomp}

\usepackage{amsfonts}
\usepackage{amsmath}
\usepackage{upgreek}     
\usepackage[T1]{fontenc} 
\usepackage{ae,aecompl}  
\newcommand{\micron}{~\ensuremath{\upmu\text{m}}}

\newcommand{\uK}{~\ensuremath{\upmu\text{K}}}

\begin{document}
\title{Measuring molecular frequencies in the 1--10\micron\ range at 11-digits accuracy}
\author{G. Insero$^{1,2}$}
\author{S. Borri$^{1,2}$}
\author{D. Calonico$^3$}
\author{P. Cancio Pastor$^1$}
\author{C. Clivati$^3$}
\author{D. D'Ambrosio$^1$}
\author{P. De Natale$^{1,2}$}
\author{M. Inguscio$^1$}
\author{F. Levi$^3$}
\author{G. Santambrogio$^{1,2,3}$}
\email{santambrogio@lens.unifi.it}
\affiliation{$^1$Istituto Nazionale di Ottica-CNR \& European Laboratory for Non-Linear Spectroscopy
  LENS, Via Nello Carrara 1, 50019 Sesto Fiorentino, Italy\\
$^2$Istituto Nazionale di Fisica Nucleare INFN, Sez. di Firenze,  Via Sansone 1, 50019 Sesto Fiorentino, Italy\\
$^3$Istituto Nazionale di Ricerca Metrologica INRIM, Strada delle Cacce 91, 10135 Torino, Italy}

\begin{abstract}
Mid infrared (MIR) photonics is a key region for molecular
physics~\cite{Borri_AdvancesinPhysicsX1p368_2016}. High-resolution
spectroscopy in the 1--10\micron\ region, though, has never been
fully tackled for the lack of widely-tunable and practical light
sources. Indeed, all solutions proposed thus far suffer from at least one of
three issues: they are feasible only in a narrow spectral range; the
power available for spectroscopy is limited; the frequency accuracy is poor.  
Here, we present a setup for high-resolution spectroscopy 
that can be applied in the whole 1--10\micron\ range by combining the
power of quantum cascade lasers (QCLs) and the accuracy achievable by
difference frequency generation using an OP-GaP crystal. The frequency
is measured against a primary frequency standard using
the Italian metrological fibre link 
network. We
demonstrate the performance of the setup by measuring a 
vibrational transition in a highly-excited metastable state of CO
around 6\micron\ with 11 digits of precision, four orders of magnitude
better than the value available in the
literature~\cite{Davies_MolPhys70p89_1990}.

\end{abstract}

\maketitle
Whereas the fractional accuracy on spectroscopic measurements on atoms
has reached the few parts in
10$^{18}$~\cite{Nicholson_NatureComm6p6896_2015}, experiments on
molecules perform worse by more than three orders of magnitude. This
is due to the richer internal structure of molecules that makes cooling and
detection more complicated than in atoms. However, the internal
structure and symmetry of molecules, and their strong intramolecular
fields can enable totally new measurements. 
Recent experiments on ThO yielded the most sensitive measurement
to date of the electron electric dipole
moment~\cite{_Science343p6168_2013}. The upper limit found in the
10$^{-29}$ $e$~cm range constrains T-violating physics at the TeV
energy scale, comparable to the energy scales explored directly at the
Large Hadron Collider. In other experiments, the
laboratory assessments of the variation of fundamental constants
based on molecular
spectroscopy~\cite{Shelkovnikov_PhysRevLett100p150801_2008,Truppe_NatureComm4p2600_2013} 
achieve a level of sensitivity similar to astronomical observations
looking back in time several billion years. Further experiments are
probing energy differences in enantiomers of chiral
species~\cite{Darquie_Chirality22p870_2010}, testing quantum
electrodynamics~\cite{Salumbides_PhysRevLett107p043005_2011}, and
searching for a fifth force~\cite{Salumbides_PhysRevD87p112008_2013}.

The MIR is a natural spectral region for high-resolution
spectroscopic studies on molecules because it coincides with
fundamental rovibrational transitions, which have strong linestrenghts
and Hz-level natural linewidths. A combination of a cold molecular
sample and state-of-the-art photonics in the MIR is the key
ingredient to catch up with the atomic physics level of precision. 

Since 2014, the development of cold molecules technology 
has accelerated. First, SrF was trapped in a three-dimensional
magneto-optical trap~\cite{Barry_Nature512p286_2014}. In the meanwhile, laser cooling
of YO~\cite{Hummon_PhysRevLett110p143001_2013} and
CaF~\cite{Zhelyazkova_PhysRevA89p053416_2014},
and optoelectrical cooling of
formaldehyde\cite{Prehn_PhysRevLett116p063005_2016} has been reported,
achieving temperatures as low as 0.5~mK at 10$^{7-8}$ molecules/cm$^3$
densities~\cite{Norrgard_PhysRevLett116p063004_2016}. Very recently,
an ammonia fountain enabling, in principle, sub-Hz linewidths has been
demonstrated~\cite{Cheng_PhysRevLett117p253201_2016}, as well as laser
cooling of CaF below the Doppler limit, to 50\uK~\cite{Truppe_arXiv1703p00580_2017}.

On the other hand, photonics in the MIR remains challenging. Coherent sources must 
feature sufficient intensity, very narrow linewidths, high frequency
stability, and an absolute frequency traceability against the primary
frequency standard. Continuous wave (CW) optical parametric oscillators (OPOs) can match these requirements
only below 5\micron\ wavelengths at the price of a difficult
operability~\cite{Ricciardi_OptLett40p4743_2015}, while 
difference frequency generation (DFG) process suffers from low emitted
powers. In the whole 1--10\micron\ range, orientation-patterned (OP) 
GaP provides a reasonable
efficiency~\cite{Insero_OptLett41p5114_2016}. 
Room-temperature QCLs cover the whole 3--25\micron\ range and
feature mW-to-W power
levels~\cite{Vitiello_OptExpress23p5167_2015}. They
have very narrow intrinsic
linewidths~\cite{Bartalini_PhysRevLett104p083904_2010}
although current noise can be very
detrimental~\cite{Borri_IEEEJQuantElec47p984_2011}. However, proper
control has yielded linewidths of several 
hundreds of
Hz~\cite{Cappelli_OptLett37p4811_2012,Hansen_OptLett40p2289_2015}. 
Transfer of the accuracy of the microwave frequency standard to the MIR has been
achieved by non-linear frequency conversion, leaving only low powers
to spectroscopic
applications~\cite{Schliesser_NaturePhot6p440_2012,Argence_NaturePhot9p456_2015,Galli_ApplPhysLett102p121117_2013}.
In addition, since the overall accuracy of the RF-MIR bridging by
optical frequency combs (OFCs) critically depends on the stability of
the OFC repetition rate, a step-change was recently determined by the
rise of frequency dissemination by optical
fibres~\cite{Clivati_OptExpress24p11865_2016}. Indeed, RF oscillators
disciplined by the Global Positioning System (GPS) can achieve
10$^{-14}$ stability and accuracy only after tens of thousands seconds integration
times. Instead, optical transfer of the metrological
performances of the primary standard allows to stabilize a
near-infrared (NIR) OFC at least at the 10$^{-14}$ level in 1~s, and
to achieve the intrinsic uncertainty of the disseminated clock with
short interaction times~\cite{Clivati_OptExpress24p11865_2016}.

All this impressive progress, however, has yet failed to trigger
significant breakthroughs in molecular physics, in particular in the 
5--10\micron\ region. Indeed, only isolated high-precision measurements below
5\micron~\cite{Galli_ApplPhysLett102p121117_2013} and 
around 10\micron\cite{Shelkovnikov_PhysRevLett100p150801_2008,Argence_NaturePhot9p456_2015} have been
reported.

\begin{figure*}[t]
\centering
\includegraphics[width=\textwidth]{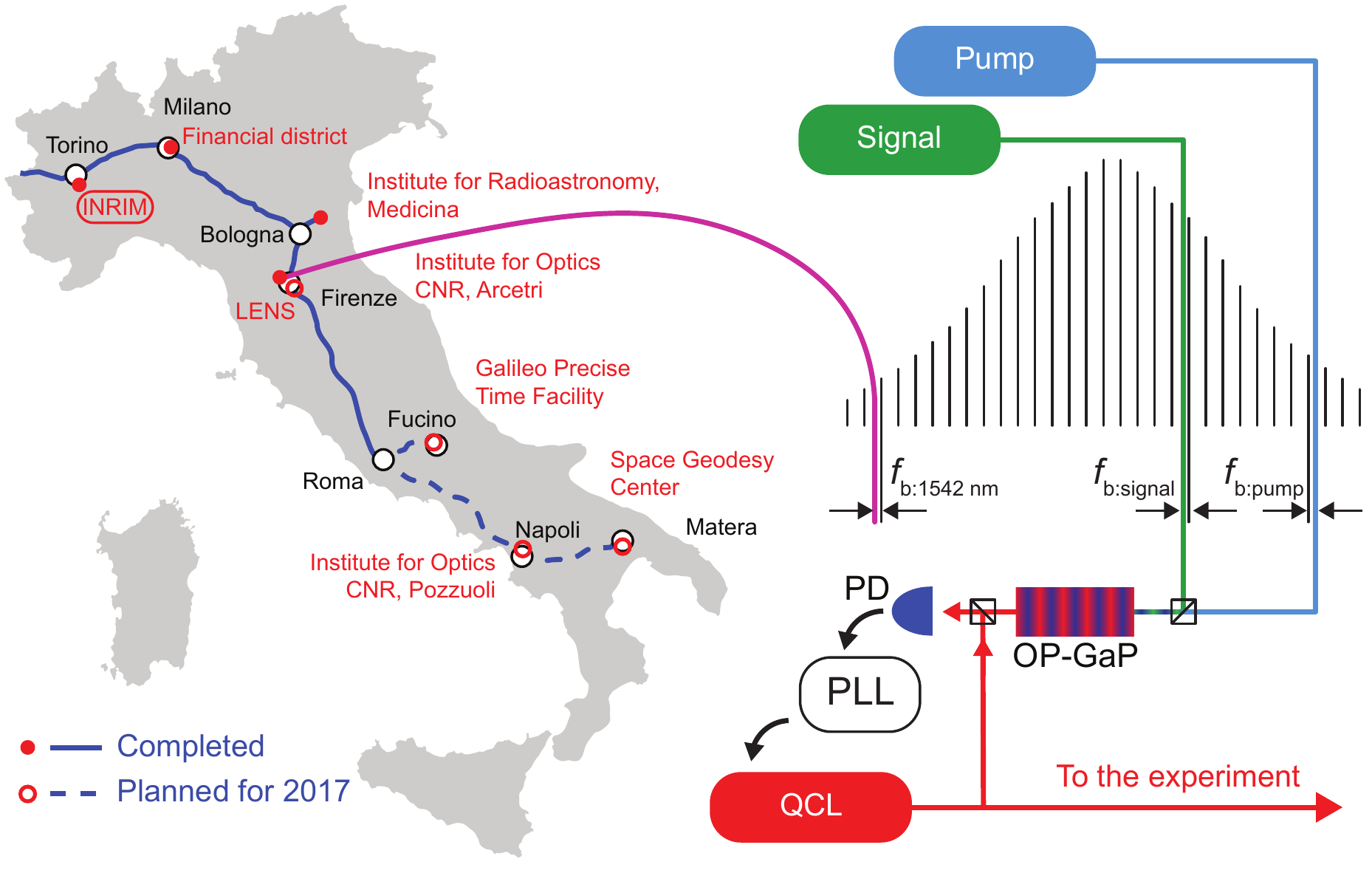}	\caption{The
left part of the figure shows the Italian fibre link network, blue
line, with network nodes in red. At LENS, the ultrastable laser at
1542~nm is used to lock the repetition rate of an OFC. The beat notes
of two NIR lasers are measured against the OFC and their frequency
difference is kept constant with a direct digital synthesis scheme to cancel
out the comb noise contribution. The two lasers are combined in an
OP-GaP crystal to generate MIR light, to which a QCL is locked.
\label{fig:frequency}}
\end{figure*}

We developed a laser technology that is suitable for the whole MIR
range between 1 and 10\micron\ with unprecedented metrological
features. We demonstrate its performance by  
measuring a vibrational transition around 6\micron\
on a highly-excited, metastable state in carbon monoxide with
kHz-level acccuracy. The experiment is done on a molecular beam with a density of
about 10$^{8}$/cm$^3$, similar to what is currently obtainable in
state-of-the-art setups for cold molecules. Our system
is based on three pillars.

\emph{First}, the frequency reference to the
primary standard is transferred by an ultrastable laser at 1542~nm
sent over the Italian fibre-link network~\cite{Calonico_ApplPhysB117p979_2014}. This network connects
several laboratories all over the national territory (details in the
Methods Section), being part of a continental metrological network
that is currently under construction. 

\emph{Second}, a CW DFG process in an OP-GaP crystal is used to bridge
the gap between the NIR and the MIR. OP-GaP is conveniently
transparent in the 1--10\micron\ range and is the most efficient
solution to date to produce light at wavelengths above 5\micron\ when
pumping at 1064~nm, where the most powerful, reliable and stable
lasers are available. We recently characterized the linear,
thermo-optic, and nonlinear properties of this
crystal~\cite{Insero_OptLett41p5114_2016}.

\emph{Third}, the MIR radiation is produced by QCLs whose output is
almost entirely used for spectroscopy and not for its frequency control.

Figure~\ref{fig:frequency} shows the main features of the
metrological chain to control MIR-QCLs. The NIR-MIR frequency gap is
bridged by CW DFG in an OP-GaP crystal
pumped by two NIR lasers.
Both NIR lasers are locked to an OFC whose repetition
rate is referenced and stabilized using the fibre link. Finally, the
DFG-MIR radiation is used to phase-lock the QCL using a beat note
signal in a HgCdTe detector. The pump laser is a Nd:YAG MOPA
system at 1064~nm with a linewidth of 1~kHz and a maximal output power
of 50~W, while the signal is provided by a diode laser delivering
about 30~mW.

\begin{figure}
\centering
\includegraphics[width=0.45\textwidth]{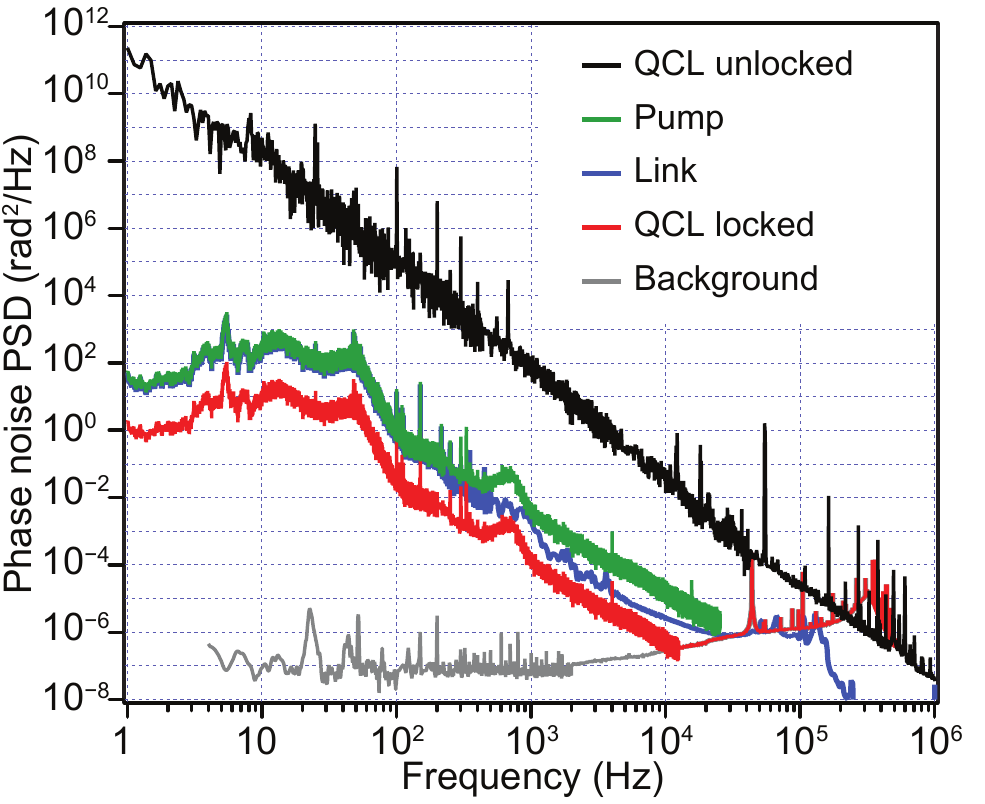}	\caption{Phase
noise of various components of the frequency locking chain. The phase
noise of the free running QCL is shown in black. The noise on the
fibre link is shown in blue. The error signal of the phase-lock loop
for the pump laser is shown in green. The expected phase noise of the
QCL when it is locked to the DFG radiation is shown in red and is
obtained by scaling the phase noise of the pump laser, following the
virtual beat note scheme.\label{fig:noise}} 
\end{figure}

The details of the locking chain are described in the Methods section. To
characterize the metrological properties of the MIR light, we analized the
phase-noise power spectral density (PNPSD) of every step of the
chain. The measurement of
the phase-noise from the optical link is shown in blue in
Fig.~\ref{fig:noise}. The pump
laser is phase-locked to the closest comb tooth on a bandwidth of
700~Hz. The residual noise of the pump laser at closed phase-lock-loop
(PPL1) (green trace in Fig.~\ref{fig:noise}) is estimated to be
limited by the comb noise within the PLL1 bandwidth. At higher
frequencies, the pump laser features its free-running noise
behaviour. The signal laser is phase-locked (PPL2) to the pump laser
through a direct digital synthesis (DDS)
scheme~\cite{Telle_ApplPhysB74p1_2002}.
Finally, the QCL is phase locked to the MIR radiation produced in the OP-GaP
crystal with a 300-kHz bandwidth PLL3.
When the phase-lock chain is operating, the phase noise of the QCL
is reduced by more than nine decades at low frequencies and by more
than two decades at 30~kHz (black and red
traces for free-running and locked, respectively).
The DFG phase noise is expected to follow that of the pump laser
scaled by a factor $(1-N_s/N_p)$, according to the DDS locking scheme,
since the noise introduced by PPL2 is negligible. A similar
consideration applies to PPL3, whose noise (grey trace) is negligible
with respect to the DFG phase noise up to few tens of kHz. Hence, we
infer that the QCL linewidth is limited by the Nd:YAG laser and that
the QCL long term stability and accuracy reflects reliably the
performance of the Cs fountain. Finally, we remark that QCL linewidth
and jitter at the kHz level can be further improved by referencing the
OFC to an ultranarrow optical-link-disciplined laser around the Nd:YAG
frequency.

\begin{figure}
\centering
\includegraphics[width=0.45\textwidth]{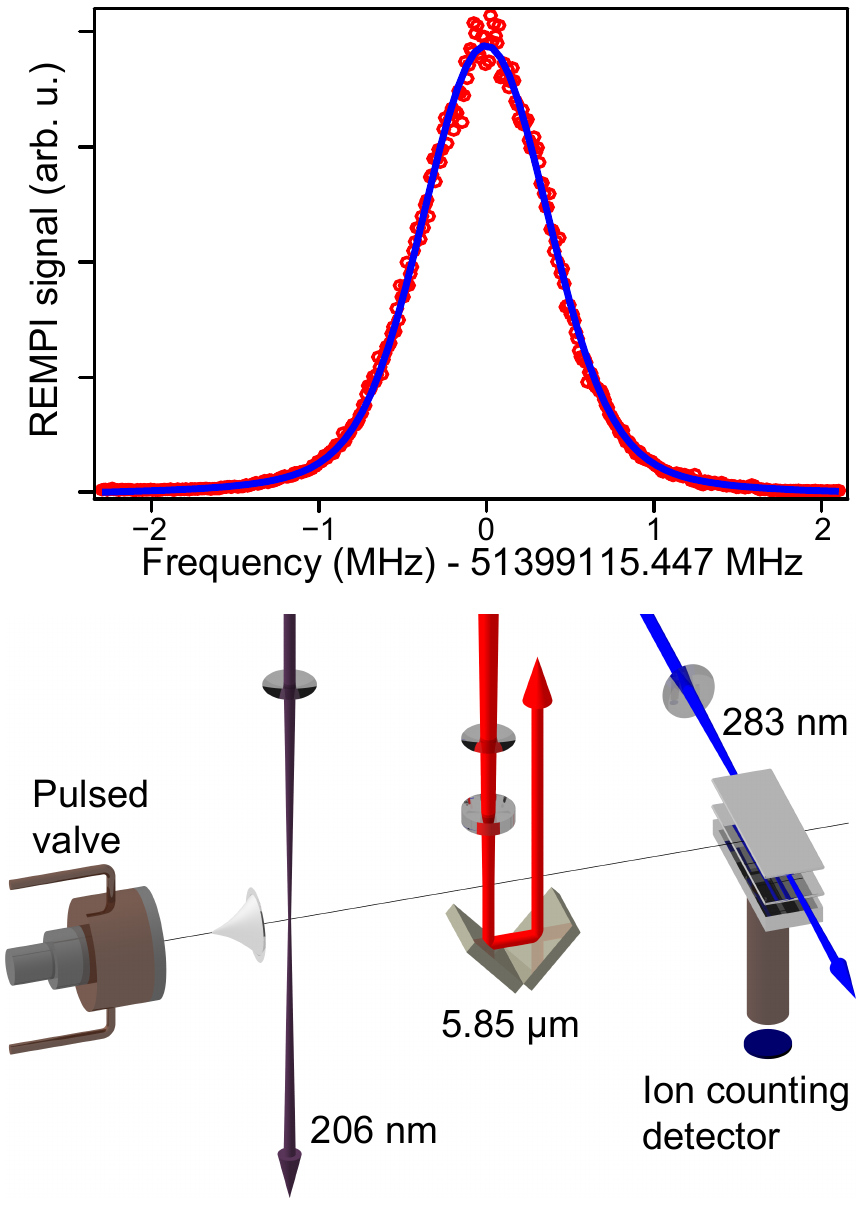}
\caption{Top: a typical vibrational absorption spectrum on the
$a^3\Pi_1$ metastable state of CO measured in about 20 minutes. The
transition $|v=0,J=1,+>\leftarrow|v=1,J=1,->$ shows a width of 900~kHz. Bottom: sketch of the molecular beam apparatus used for the
measurement. The beam is generated by a pulsed valve operated at
10~Hz. CO molecules are skimmed, excited into the metastable state 
by a UV laser at 206~nm, interact with the mid IR laser, and are finally
detected by resonance-enhanced multiphoton ionization. Ions are
collected on a microchannel plates detector.
\label{fig:experiment}}  
\end{figure}

To prove the versatility of our setup, we measured a vibrational
transition on a slow (318~m/s) molecular beam of highly-excited, metastable CO
molecules, in a triple resonance experiment. Over 90\% of the QCL power can be
coupled to the molecular beam, i.e.\ about 10~mW. For a typical transition
dipole moment of 0.05~debye, this power corresponds to a Rabi
frequency of some MHz for a beam waist of 1~mm, which are comfortable
numbers for many experiments. The metastable $a^3\Pi_1$, $v=0$, $J=1$,
$+$ state is prepared with a
pulsed laser at 206~nm (1 mJ, 150~MHz bandwidth), whereas the
molecules are detected via multiphoton resonance-enhanced ionization
from the $v=1$, $J=1$, $-$ state with a $1+1$ process at 283~nm, after the
interaction with the MIR light. The experiment runs at 10~Hz. The
stability and robustness of this triple resonance setup allows for
rapid scans yielding high signal-to-noise ratios. A 25-minutes
scan yields a Doppler-broadened absorption profile with a FWHM of
900~kHz, as the one shown in Fig.~\ref{fig:experiment}, which is
already about two orders of magnitude narrower
than the best value reported in the literature
($\pm$90~MHz)~\cite{Davies_MolPhys70p89_1990}. After 9 hours
averaging, we could determine the line center of this vibrational
transition on the metastable excited state with an uncertainty of
3~kHz as 51399115447 kHz, with and improvement of more than 4 orders
of magnitude~\cite{Davies_MolPhys70p89_1990}. The relative uncertainty
of $6\times 10^{-11}$ is mainly due to systematic effects to which the
QCL frequency uncertainty and linewidth contributes with $2\times 10^{-14}$ at worst.

We described the accurate frequency traceability from 1 to 10\micron.
The range below 3\micron, which is currently not covered by QCLs, is
somehow trivial for the availability of various solid state lasers and
very efficient non-linear crystals. Thus, our setup extends
high-resolution, accurate frequency measurements to 
the whole molecular fingerprint region. Frequency dissemination over
fibre network allows, for instance, to synchronize
of radio telescopes, realizing a thousand-km wide interferometer, whereas
synchronization of spectroscopy laboratories allows
for simultaneous monitoring of atomic and molecular transitions
with unprecedented accuracy that can shed light, for instance, on
topological defect dark
matter~\cite{Stadnik_ModPhysLettA29p1440007_2014}.
Moreover, the fibre network together with state-of-the-art MIR
photonics and innovative techniques of molecular beam manipulation
brings atomic-level accuracies within reach. In particular, the
10$^{-16}$ accuracy allowed by the fibre link can become the standard
accuracy for measurements in the MIR ($\sim 10^{13}$~Hz), when
molecular cooling techniques will allow for 1-second interaction
times. This can disclose new perspectives in fundamental sciences,
leading to the replacement of large-scale infrastructures with
tabletop setups probing physics at sub-eV energies.

\section{Methods}

\subsection{The frequency locking chain}

The fibre link network carries the light of
an ultrastable laser that is kept at a frequency of
194399996000000.00(4) Hz (i.e.\ around 1542~nm), where the uncertainty
is directly related to that of the Cs fountain. The laser frequency is referenced to
the Italian timescale, generated at the Italian Metrological Institute
(INRIM) in Torino. This signal reaches several relevant locations on
the national territory: the 
Institute for Radioastronomy in Medicina (535~km from INRIM), the
European Laboratory for Nonlinear Spectroscopy, LENS, in 
Sesto Fiorentino (642~km),
Rome (994~km), and the Italian-French border (150~km). In addition, a
timestamp signal referenced to the Italian timescale reaches the
financial district in Milano (279~km). During 2017, the
fibre network will be extended to the National Institute for Optics
INO in Arcetri (662~km), to CNR labs in Pozzuoli
(1306~km), to the Galileo Precise
Time Facility in Fucino (1134 km), to the Space Geodesy Center in Matera
(1684~km), and finally to the French metrological institute LNE-SYRTE in
Paris. 

A series 
of bidirectional erbium-doped fibre amplifiers compensate for the
almost 200~dB optical losses of the link between INRIM and LENS~\cite{Calonico_ApplPhysB117p979_2014}.
The laser at 1542~nm has a
stability of $1\times 10^{-14}$ at 1~s and an accuracy of 
$2\times 10^{-16}$; these performances have been assessed by
measuring the absolute frequency of the 1542-nm laser on two
independent optical combs referenced to the same H-maser.
The optical link does
not affect the uncertainty of the delivered signal within parts in
10$^{19}$~\cite{Calonico_ApplPhysB117p979_2014}. However, the 
signal delivered over the fibre is affected by a large amount of phase noise in
the short term (<5~ms) due to the limited bandwidth consequent to the photon
round-trip time in the fibre~\cite{Clivati_OptExpress24p11865_2016}.

At LENS a diode laser is phase-locked to the incoming
radiation, replicating the stability of the link laser and boosting
the optical power at a suitable level for referencing an OFC. 
The repetition rate $f_\text{rep}$ of a commercial OFC is then
phase-locked to the 1542-nm light using an intra-cavity
electro-optic modulator, with a bandwidth of approximately
300~kHz; the lock frequency is 
properly chosen so that the repetition rate is 100~MHz when the
incoming radiation is at the nominal frequency value. The
carrier-envelope offset frequency $f_0$ is stabilized to an
RF frequency standard. The phase-noise of the optical comb is then
dominated by the residual noise of the stabilized link,
which is limited by the fiber length \cite{Williams_JOptSocAmB25p1284_2008}.

A Nd:YAG MOPA system at 1064~nm with a linewidth of
1~kHz (pump) and a diode laser at 1301~nm (signal) are locked to the
OFC. The beat notes 
of both lasers to the closest teeth of the OFC are detected with two
fast InGaAs photodiodes with a minimum signal-to-noise ratio of
25~dB on a 100~kHz bandwidth. We refer to these beat notes as
$f_\text{b:pump}$ and $f_\text{b:signal}$, and we refer to the
absolute frequencies of the two lasers as $\nu_\text{pump}$ and
$\nu_\text{signal}$.
Figure \ref{fig:noise} shows
the phase noise of the beat note $f_\text{b:pump}$ between the pump and
the closest comb teeth. The link noise is derived from
independent measurements \cite{Clivati_OptExpress24p11865_2016}, and
re-scaled to the respective spectral region.
The difference frequency 
$\text{DF}=\nu_\text{pump} - \nu_\text{signal}$ is stabilized by an
indirect lock of the signal laser to the pump. This is done via the
DDS scheme~\cite{Telle_ApplPhysB74p1_2002},
where the OFC bridges the frequency gap between the two lasers but its
noise contribution is rejected by proper processing of the
$f_\text{b:pump}$, $f_\text{b:signal}$, and $f_0$.

To implement the DDS, $f_0$ is subtracted from
each beat note using analog mixers, to produce
$\overline{f}_\text{b:pump}$ and $\overline{f}_\text{b:signal}$. Thus, the absolute
frequencies of the two lasers can be written as $\nu_\text{pump}=N_\text{pump}
f_\text{rep} +\overline{f}_\text{b:pump}$ and $\nu_\text{signal}= N_\text{signal}
f_\text{rep} + \overline{f}_\text{b:signal}$, where $N_\text{pump}=2816363$ and
$N_\text{signal}=2302371$ are the numbers of the comb's teeth to which the two lasers
are beaten. The signal
$\overline{f}_\text{b:signal}-(N_\text{signal}/N_\text{pump})\overline{f}_\text{b:pump}$
is generated with a 14-bit Direct Digital Synthesizer (DDS) and a
mixer, and is stabilized to a RF frequency reference using a PLL
which feeds back the signal laser on a bandwidth exceeding 200~kHz. The finite
number of digits in the DDS generates a bias of about 1~Hz in the frequency
difference. However, this is calculated and corrected in the final
results.

\subsection{Error analysis for the measurement of the vibrational transition}

To minimize the systematic Doppler shift due to the imperfect
perpendicularity  between the MIR laser beam and the molecular beam,
the laser light is retro-reflected by a corner cube. Each of the two
anti-parallel laser beams induces a Doppler shift on the transition that
is equal in magnitude to the other but opposite in sign. Thus, a
deviation from the perpendicularity condition manifests itself as a
symmetric splitting of the absorption line. The alignment procedure
consists in scanning the angle between the laser beam and the
molecular beam while recording the magnitude of the splitting;
finally, the position of minimal splitting is chosen. The remaining systematic
uncertainty is due to imperfect parallelity of the counter-propagating MIR
beams. This has been measured and it is better than 10$^{-4}$~rad,
which corresponds to a final uncertainty on the transition frequency of
2.6~kHz.

We estimate the Stark and second-order Zeeman shifts due to stray
fields to be lower than 10~Hz. As stated above, we correct for the
bias induced by the finite number of digits of the DDS that does not
allow to set the perfect value for the DDS. However, the less
significant digit of the device introduces a 
systematic uncertainty that we cannot compensate, which is smaller than
1~Hz. Since the molecular population is first prepared and then detected with
focused ns lasers at precisely-known times and positions, we
calculate the speed of the molecular beam as 318.5$\pm2$~m/s. Thus,
the second-order Doppler shift, $+29.4\pm0.4$~Hz, is subtracted from
the measured frequency.
The remaining systematic uncertainty are due to the Cs fountain
standard accuracy~\cite{Levi_Metrologia51p270_2014} and to fibre link
phase slips~\cite{Clivati_OptExpress24p11865_2016}, both smaller than
100~mHz. Therefore, the \emph{total systematic uncertainty}
is estimated as 2.6~kHz.

The Zeeman shift of the $\Delta M=\pm 1$ is of the order of
500~kHz/Gauss, whereas the $\Delta M= 0$ are shifted by about
10~kHz/Gauss. By canceling the Zeeman shift on the $\Delta M=\pm 1$
transitions, we estimate that the residual Zeeman shift on the $\Delta
M= 0$ transition is smaller than 1~kHz. This contributes to the
statistical uncertainty, together with the uncertainty in the alignment of the
beams, whose procedure is described above, and with the fluctuations
in the population of metastable CO molecules. The distribution of the
experimental measurements has a Gaussian shape with a standard
deviation of 7.9~kHz. Our set of 22 scans yields a \emph{total
  statistical uncertainty} of 1.7~kHz.

\bibliography{bib}
\bibliographystyle{gams-notit-nonumb}

\section{Acknowledgements}
This work was supported by INFN under the SUPREMO project, and by
EMPIR-15SIB05-OFTEN, which has received funding from the EMPIR
programme co-financed by the Participating States and from the
European Union't Horizon 2020 research and innovation programme. 
The authors gratefully acknowledge P. G. Schunemann and BAE Systems,
Inc.\ for providing the OP-GaP crystal; A. Mura and M. Frittelli for their
assistance in the fibre link operation; and M. De Pas for the design
and the realisation of the locking electronics at LENS. 
\end{document}